\documentclass[journal]{IEEEtran}
\usepackage[center]{caption}
\usepackage{subcaption, cite, stfloats, multicol}

\ifCLASSINFOpdf
   \usepackage[pdftex]{graphicx}
\else
\fi
\usepackage{amsmath}

\usepackage{color, soul}
\usepackage{tabularx}
\usepackage{lipsum}

\soulregister\cite7
\soulregister\ref7
\soulregister\eqref7
\soulregister\pageref7
\sethlcolor{yellow}

\makeatletter
\def\SOUL@hlpreamble{%
   \setul{\dp\strutbox}{\dimexpr\ht\strutbox+\dp\strutbox\relax}%
   \let\SOUL@stcolor\SOUL@hlcolor
   \SOUL@stpreamble
}

\begin{document}
%
% paper title
% Titles are generally capitalized except for words such as a, an, and, as,
% at, but, by, for, in, nor, of, on, or, the, to and up, which are usually
% not capitalized unless they are the first or last word of the title.
% Linebreaks \\ can be used within to get better formatting as desired.
% Do not put math or special symbols in the title.
%\title{Uncertainty aware deep neural network framework for guided wave based damage localization}
\title{Uncertainty Aware Deep Neural Network for Multistatic Localization with Application to Ultrasonic Structural Health Monitoring}
%
%
% author names and IEEE memberships
% note positions of commas and nonbreaking spaces ( ~ ) LaTeX will not break
% a structure at a ~ so this keeps an author's name from being broken across
% two lines.
% use \thanks{} to gain access to the first footnote area
% a separate \thanks must be used for each paragraph as LaTeX2e's \thanks
% was not built to handle multiple paragraphs
%

\author{Ishan D. Khurjekar, Joel B. Harley,~\IEEEmembership{Member,~IEEE}
       % <-this % stops a space
%\thanks{M. Shell was with the Department
%of Electrical and Computer Engineering, Georgia Institute of Technology, Atlanta,
%GA, 30332 USA e-mail: (see http://www.michaelshell.org/contact.html).}% <-this % stops a space
\thanks{I. Khurjekar and J.B. Harley are with the Department of Electrical and Computer Engineering at the University of Florida.}% <-this % stops a space
}

% note the % following the last \IEEEmembership and also \thanks - 
% these prevent an unwanted space from occurring between the last author name
% and the end of the author line. i.e., if you had this:
% 
% \author{....lastname \thanks{...} \thanks{...} }
%                     ^------------^------------^----Do not want these spaces!
%
% a space would be appended to the last name and could cause every name on that
% line to be shifted left slightly. This is one of those "LaTeX things". For
% instance, "\textbf{A} \textbf{B}" will typeset as "A B" not "AB". To get
% "AB" then you have to do: "\textbf{A}\textbf{B}"
% \thanks is no different in this regard, so shield the last } of each \thanks
% that ends a line with a % and do not let a space in before the next \thanks.
% Spaces after \IEEEmembership other than the last one are OK (and needed) as
% you are supposed to have spaces between the names. For what it is worth,
% this is a minor point as most people would not even notice if the said evil
% space somehow managed to creep in.

% The paper headers
\markboth{This work has been submitted to the IEEE for possible publication}%
{Khurjekar and Harley: Uncertainty-aware Deep Neural Network Localization}

% If you want to put a publisher's ID mark on the page you can do it like
% this:
%\IEEEpubid{0000--0000/00\$00.00~\copyright~2015 IEEE}
% Remember, if you use this you must call \IEEEpubidadjcol in the second
% column for its text to clear the IEEEpubid mark.

% use for special paper notices
%\IEEEspecialpapernotice{(Invited Paper)}

% make the title area
\maketitle

% As a general rule, do not put math, special symbols or citations
% in the abstract or keywords.
\begin{abstract}
Guided ultrasonic wave localization uses spatially distributed multistatic sensor arrays and generalized beamforming strategies to detect and locate damage across a structure. The propagation channel is often very complex. Methods can compare data with models of wave propagation to locate damage.  Yet, environmental uncertainty (e.g., temperature or stress variations) often degrade accuracies. This paper uses an uncertainty-aware deep neural network framework to learn robust localization models and represent uncertainty. We use mixture density networks to generate damage location distributions based on training data uncertainty. This is in contrast with most localization methods, which output point estimates. We compare our approach with matched field processing (MFP), a generalized beamforming framework.  The proposed approach achieves a localization error of $0.0625$~m as compared to $0.1425$~m with MFP when data has environmental uncertainty and noise. We also show that the predictive uncertainty scales as environmental uncertainty increases to provide a statistically meaningful metric for assessing localization accuracy.
\end{abstract}

% Note that keywords are not normally used for peerreview papers.

% For peer review papers, you can put extra information on the cover
% page as needed:
% \ifCLASSOPTIONpeerreview
% \begin{center} \bfseries EDICS Category: 3-BBND \end{center}
% \fi
%
% For peerreview papers, this IEEEtran command inserts a page break and
% creates the second title. It will be ignored for other modes.
\IEEEpeerreviewmaketitle

\section{Introduction}
% The very first letter is a 2 line initial drop letter followed
% by the rest of the first word in caps.
% 
% form to use if the first word consists of a single letter:
% \IEEEPARstart{A}{demo} file is ....
% 
% form to use if you need the single drop letter followed by
% normal text (unknown if ever used by the IEEE):
% \IEEEPARstart{A}{}demo file is ....
% 
% Some journals put the first two words in caps:
% \IEEEPARstart{T}{his demo} file is ....
% 
% Here we have the typical use of a "T" for an initial drop letter
% and "HIS" in caps to complete the first word.

\IEEEPARstart {L} {ocalization} is an ubiquitous problem and has been researched across a wide range of fields. It has been studied for applications such as, underwater acoustics \cite{underwatersonar}, seismology \cite{sidorovich1998two}, sensor networks \cite{safavi2017localization}, and radar \cite{AdaptiveRadar}. Extensive theory-based localization methods have been developed by signal processing researchers. These include model-based methods \cite{alpay2000model}, beamforming \cite{BeamformingLocalization}, subspace-based methods \cite{zuo2018subspace}, and multi-lateration \cite{le2016closed}. In this work, we consider the application of localization in ultrasonic structural health monitoring (SHM). 

In recent years, there has been an increased demand to maintain new and existing physical infrastructures, including buildings, aircraft, and oil rigs. This demand has increased interests in SHM. \cite{ko2005technology}. SHM engineers create technology to detect, locate, and characterize damage in structures. In SHM, guided ultrasonic wave (GUW) systems are extensively studied for damage localization \cite{rose2009successes, mitra2016guided}. GUWs have several attractive properties, such as a long scanning range, high penetrability, and greater damage sensitivity. These systems use ultrasonic sensors to transmit and / or receive waves across the structure. The sensors are often spatially dispersed to monitor large areas. Transmitters are excited at desired frequencies and the received signal is analyzed to locate the damage.

Damage localization is often solved as an inverse problem \cite{friswell2007damage}, where the damage is a point scatterer of incident waves. Methods such as MFP (an extension of the matched filter \cite{giannakis1990signal}) is a popular model-based method for damage localization \cite{harley2014data}. In MFP, a physical model for wave propagation in ideal conditions is adopted. The model (i.e., a multi-frequency waveform) is defined for every point in a spatial grid over which the damage is located. The model at every point is then compared with experimental data. The grid point with the maximum correlation is the estimated damage location.

As damage localization is a critical task in SHM \cite{farrar2007introduction}, analyzing sources of uncertainty is important. Many structures, such as aircrafts, experience extreme conditions. This alters the nature of the collected SHM data and necessitates robust signal processing to identify and locate damage in real-time prior to catastrophic failures and economic losses. As a result, there has been research on uncertainty quantification for SHM \cite{lorenzoni2016uncertainty, sankararaman2011uncertainty}. 

\indent Within the realm of source localization, researchers have studied the effect of uncertainties in sensor locations on the localization performance \cite{EllipticLocalization, RecLocnError}. Yet, for commonly used damage localization techniques in SHM systems, such as model-based MFP, effects of environmental variations is a significant challenge \cite{feuillade1989environmental}, \cite{StochasticMFP}. For example, wave velocity has a strong temperature dependence \cite{Temperature, raghavan2008effects}. This leads to deviations from the physical model of wave propagation. The characterization of these external uncertainties is a challenging research problem \cite{eybpoosh2014investigation}, which limits the performance of model-based localization algorithms. Hence, it is of prime importance to account for these environmental uncertainties when building a SHM system \cite{harley2012scale}.

\indent Data-driven methods, such as machine learning, are another approach for localization. Deep neural networks (DNN) have produced remarkable results on a variety of signal processing problems \cite{InterferenceManagement, MIMODeepLearning, ResourceAlloc}. DNN based localization algorithms are found in various applications, such as underwater acoustics \cite{niu2017source}, sound source localization \cite{yalta2017sound}, and object localization \cite{tompson2015efficient}. Neural networks learn non-linear representations from input to output \cite{bengio2012deep}. This makes them ideal for SHM localization, which involve complex dependencies between inputs and outputs \cite{harley2017managing}. Yet end-to-end learning of desired parameters, as is common in data-driven paradigms, is not well-suited for parameter estimation problems where uncertainty has to be accounted for suitably. This motivates us to incorporate and represent uncertainty in damage localization. 

\begin{figure}[t]
    \centering
    \includegraphics[width = 9.0 cm]{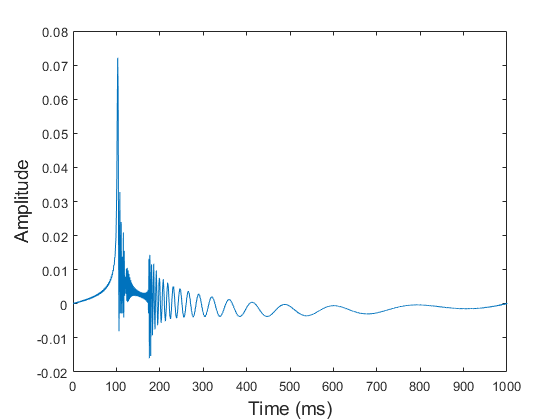}
  \caption{Illustration of a Lamb wave impulse response, as defined by \eqref{lambwave} for a distance $r$ = 0.5407 m. The function $\kappa_n(\omega)$ is defined by the solution of the Rayleigh-Lamb equation \cite{tian2014lamb} for a 1 m by 1 m aluminum plate.}
    \label{wave}
\end{figure}

\begin{figure*}[!ht]
  \begin{subfigure}[h]{0.5\textwidth}
  \centering
  %\captionsetup{justification=centering}
    \includegraphics[width=0.9\textwidth]{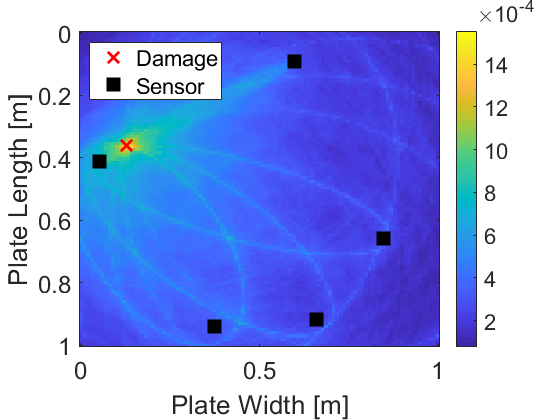}
    \caption{}
    \label{mfp_ideal}
  \end{subfigure}
  \hfill
  \begin{subfigure}[!hb]{0.5\textwidth}
  \centering
  %\captionsetup{justification=centering}
    \includegraphics[width=0.9\textwidth]{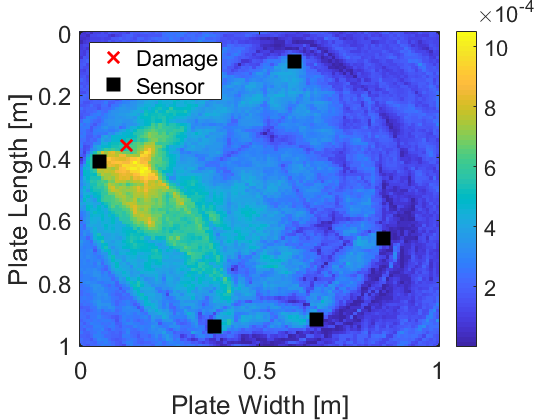}
    \caption{}
    \label{mfp_velvar}
  \end{subfigure}
  \caption{Ambiguity heatmap produced by MFP in presence of: (a) no uncertainty and (b) uncertainty in velocity (wavenumber distortion = 0.15) + 5dB noise} 
\end{figure*}

We propose a DNN based framework for damage localization in presence of uncertainty. We model environmental uncertainty as randomness in the wave velocity / wavenumber. We model sensor noise as additive white Gaussian noise. We model the damage location as a multi-modal probability distribution. The distributional parameters are obtained as outputs from the DNN. The variance estimate of the output distribution represents predictive uncertainty. By using multiple mixture components in the distribution, this framework can identify multiple damage locations. 

We train and validate the DNN model using simulated data. The DNN framework is compared with MFP for varying levels of uncertainty. The performance of the two frameworks is also compared with varying numbers of damages. The proposed framework demonstrates robust localization that quantifies predictive uncertainty in the process. Our approach demonstrates a localization error of $0.0625$~m as compared to $0.1425$~m with MFP in presence of environmental uncertainty and $5$~dB noise.  In addition, we show that the predictive uncertainty scales as environmental uncertainty increases to provide a statistically meaningful metric for assessing localization accuracy.

\section{\label{sec:3}Methodology: \\Uncertainty aware damage localization}

\subsection{Localization Input Uncertainties}
Lamb waves are a specific type of guided waves found in isotropic plates and used for damage localization \cite{alleyne1992interaction}. The Lamb wave frequency response is defined by
\begin{equation}\label{lambwave} 
    X(\omega,r) = \sum_{n}\sqrt \frac{1}{\kappa_{n}(\omega)r}e^{-j\kappa_{n}(\omega)r} 
\end{equation}
where $X(\omega, r)$ is the frequency domain representation of the signal and is modeled as the summation across $n$ wave modes. The function $\kappa_{n}(\omega)$ is the frequency and mode dependant wavenumber (known as the dispersion relation) and $r$ is the distance travelled by the wave.  Fig.~\ref{wave} illustrates the impulse response of \eqref{lambwave} with $n=2$~modes (the zeroth symmetric mode and zeroth asymmetric mode).

\indent In physical / engineering systems, researchers have formalized uncertainty into two types: aleatoric and epistemic \cite{chowdhary_dupuis_2013}. Aleatoric uncertainty varies with each experiment. In the damage localization scenario, random sensor noise is the aleatoric uncertainty. We represent random sensor noise as additive white Gaussian noise. Epistemic uncertainty is the systematic uncertainty or lack of underlying knowledge in an experiment. Uncertainty because of external/environmental factors, such as temperature, represents epistemic uncertainty. As we saw earlier, variations in temperature (as well as other environmental factors) affect wave velocity \cite{raghavan2008effects}. The dispersion relation for a wave relates group velocity to the frequency-dependent wavenumber according to \cite{weaver1981dispersion}
\begin{equation}\label{wavenumber}
    \nu_{g} = \frac{\partial \omega}{\partial \kappa} \; .
\end{equation}
From \eqref{lambwave} and \eqref{wavenumber}, we can simulate uncertainty in wave velocity by adding uncertainty to $\kappa_{n}(\omega$).

\indent Fig.~\ref{mfp_ideal} and~\ref{mfp_velvar} illustrates how uncertainty negatively affects MFP. The maximum value of the heat-maps correspond to the predicted damage location. As uncertainty increases, MFP is less accurate and less localized. The maximum value also reduces with uncertainty. Researchers have proposed data-driven methods for localization with uncertainty \cite{harley2014data}, but these techniques can require an impractical amount of calibration data and do not provide predictive uncertainty. As discussed previously, DNN based frameworks can leverage the data available from observations and learn robust representations in a localization setup. This motivates us to build an uncertainty-aware DNN framework for damage localization.

\subsection{\label{sec:3.1} Challenges with DNN-based Localization}
Consider the following localization setup. Let $\mathbf{x}$ be the input wave data from an array of sensors and let $\mathbf{y}$ be the damage location (ground truth). Let $\widehat{f}$, be a non-linear function from input to output, learnt by a DNN during training. Let $\mathbf{\widehat{y}}$ denote the DNN prediction for the damage location
\begin{equation}\label{dnn_pred}
    \mathbf{\widehat{y}} = \widehat{f}(\mathbf{x})
\end{equation}
For regression tasks (localization is a regression task), the traditional loss function is the mean squared error (MSE) between prediction of the DNN and the ground truth
\begin{equation}\label{mse_exp}
    \textrm{MSE} = E[(\mathbf{y} - \widehat{f}(\mathbf{x}))^{2}] \; .
\end{equation}
Note that MFP can also be expressed as an approach which minimizes the MSE between the model and data \cite{harley2014data}.

\begin{figure}[b!]
    \includegraphics[width= 9.0 cm]{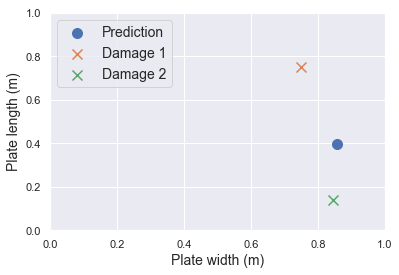}
    \caption{Damage localization example with a traditional DNN}
    \label{end2end}
\end{figure}

\indent This DNN based localization setup has a number of challenges. First, the mapping $\widehat{f}$ learned is usually an arbitrary function with no statistical information. By training such a network with an objective function of squared error, we obtain the approximate conditional mean and variance of the prediction through Monte-Carlo simulations \cite{bishop1994mixture}. Yet these two statistics are not enough to uniquely identify the output probability distribution. To represent uncertainty at the output, an appropriate probability distribution must be inferred.

\indent Second, the output of an optimization objective in \eqref{mse_exp} is a point estimate. A critical drawback with a model that generates point estimates is the scenario when we have a one-to-many mapping from input to output. Such a situation is shown in Fig.~\ref{end2end}. The neural network approximately identifies the mean of the true damage locations. This is a typical localization result of a standard feed-forward DNN (changing the DNN hyperparameters does not significantly change the result) when multiple damage locations are present.  Hence, a standard DNN is not appropriate for multi-damage localization tasks.

\subsection{Addressing Challenges using Mixture Density Networks}

To address these deficiencies in traditional neural networks, we first add a source of variability at the input stage. The predictive uncertainty (uncertainty that propagates from input data to the output) is then represented by mixture density networks, proposed in \cite{bishop1994mixture}. A mixture density network (MDN) models the conditional output probability distribution as a specific class of mixture model. In an MDN, instead of optimizing the MSE as in \eqref{mse_exp}, the conditional output likelihood is maximized. This makes MDN an attractive option to infer relevant probability distribution parameters. Second-order statistics, such as variance, can represent predictive uncertainty.

There has been renewed interest in MDN's in the context of representing uncertainty in systems \cite{choi2018uncertainty} and for applications that involve multi-modal outputs \cite{zen2014deep, wang2016gating, wang2017autoregressive}. With a MDN, we obtain two complementary advantages. First, we have the flexibility of learning non-linear representations from input to output by virtue of a deep learning based model. Second, we also have statistical information about the output distribution by virtue of likelihood optimization. In contrast, many uncertainty estimation techniques in deep learning take a Bayesian approach \cite{bernardo2009bayesian,neal2012bayesian, gal2016dropout}. These techniques typically choose a prior distribution and develop a scheme for approximating the posterior distribution. The performance is often limited by computational constraints. With an MDN, the parametric estimation is free from the computational complexities. 

There exist a number of multi-modal distributions that mixture density networks can use \cite{mclachlan2004finite}. We choose a GMM since a GMM with the appropriate number of components is capable of representing an arbitrary function within a finite error threshold \cite{kostantinos2000gaussian}. Formally, we model the probability density of our output $y_i$ conditioned on the input $x_i$ as a multi-variate GMM with $K$ components, given by  
\begin{equation}\label{prob_eqn}
    p(y_{i} | x_{i}) = \sum_{i=1}^{K}\pi_{i} \mathcal{N}(\mu_{i},\Sigma_{i})
\end{equation}
with the Gaussian mixture component $i$ given by $\mathcal{N}(\mu_{i},\Sigma_{i})$, and mixing coefficient $\pi_{i}$. The mixing coefficients must satisfy
\begin{equation}\label{mix_prob}
    \sum_{i = 1}^{K} \pi_{i} = 1
\end{equation}

\begin{figure*}[!th]
\centering
    \includegraphics[width= 18.0 cm]{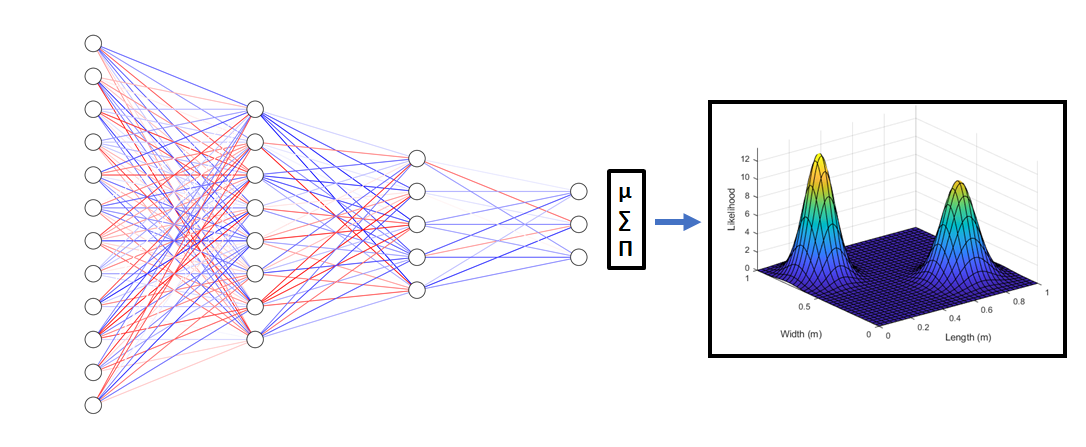}
    \caption{Mixture density network}
    \label{mdn}
\end{figure*}

In an MDN, we have the traditional architecture of a neural network with hidden layers. The last layer (referred to as the mixture density network layer) outputs a vector of length $(2d+1) k$, where $k$ is the number of mixture components and $d$ is the dimension of output space (i.e., 2 in our case). This layer outputs elements corresponding to the parameters of the mixture distribution ($dk$ elements corresponding to the mean estimates for all mixture components, $dk$ elements corresponding to variance estimates for all mixture components and $k$ elements corresponding to mixture probabilities for all mixture components). Let us denote the output vector by $\mathbf{z}$ as follows
\begin{equation}\label{output}
    \mathbf{z} = [\mathbf{z}^{\mu} ; \mathbf{z}^{\sigma} ; \mathbf{z}^{\pi}]
\end{equation}

Eq. (\ref{mu_exp},~\ref{sigma_exp},~\ref{pi_exp}) are the equations for the component means, component variances, and mixture probabilities, respectively.
\begin{equation}\label{mu_exp}
    \mathbf{\mu_{i}} = \mathbf{z}^{\mu}_{i} \qquad\forall i = 1, 2 \ldots  k
\end{equation}
\begin{equation}\label{sigma_exp}
    \mathbf{\Sigma_{i}} = e^{\mathbf{z}^{\sigma}_{i}} \qquad\forall i = 1, 2 \ldots  k
\end{equation}
\begin{equation}\label{pi_exp}
    \mathbf{\pi_{i}} = \frac{e^{\mathbf{z}^{\pi}_{i}}}{1 + e^{\mathbf{z}^{pi}_{i}}}  \qquad\forall i = 1, 2 \ldots  k
\end{equation}

Since the only restriction on the component means is that they have to be real-valued, the relation is as given in \eqref{mu_exp}. Since the component variances have to be positive, we choose a monotonically increasing function (exponential) and we have the relation as in \eqref{sigma_exp}. For the mixing coefficients to satisfy the condition in \eqref{mix_prob}, we use a soft-max function in \eqref{pi_exp}. Together, the component means, variances, and mixing probabilities specify a unique GMM. An illustration of an MDN with two mixture components at the output is shown in Fig \ref{mdn}. 

In an MDN, the conditional output likelihood as given in (5) is maximized. The distribution parameters to be used for the likelihood calculation are obtained from the preceding three equations. Typically, optimization in the deep learning paradigm is a batched process. The batch-data likelihood given by the joint conditional distribution of all output samples in a batch has to be calculated in the optimization process. Consider we have $n$ samples in one batch. Assuming independent and identically distributed data (i.i.d), the joint distribution can be written out as the product of individual output probabilities
\begin{equation}
    \mathcal{L} \propto \prod_{i}^{n}p(y_{i} | x_{i})
\end{equation}
For computational ease, the log-likelihood is optimized instead of likelihood, given by 
\begin{equation}
    l(y_{i} | x_{i}) \equiv \log (\mathcal{L}) =  \sum_{i}^{n} \log ( p(y_{i} | x_{i})) \; .
\end{equation}
The distributional parameters of the GMM can then be learned by maximizing $l(y_{i} | x_{i})$. Note that incorporating the likelihood as a loss function is non-trivial and there are computational issues that have to be considered, such as unstable learning and collapsing modes. We refer to \cite{mdn} for practical implementations of an MDN.

\subsection{Damage Localization Mixture Density Network}

\begin{figure*}[!th]
\centering
    \includegraphics[width= 16.0 cm, trim=1cm 1cm 1cm 1cm, clip]{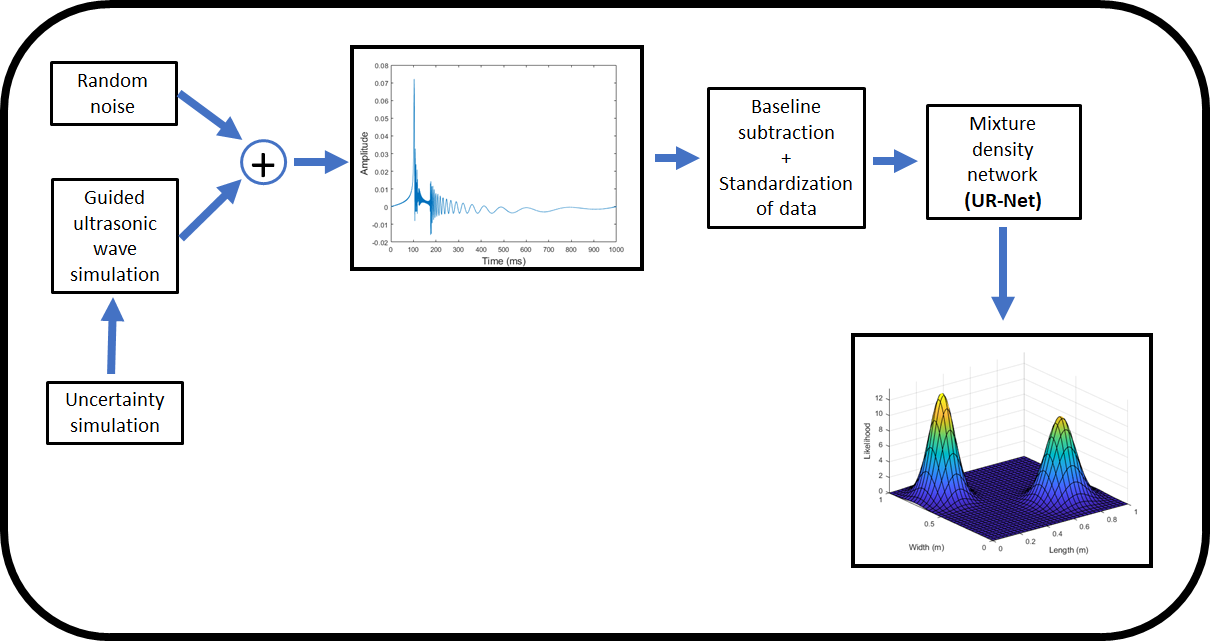}
    \caption{UR-Net Framework}
    \label{dnn_framework}
\end{figure*}

Due to its capabilities, we use an MDN trained on wave data to obtain a multi-modal distribution to estimate damage location(s) and represent predictive uncertainty. We denote our localization neural network as UR-Net (short for uncertainty representation network), shown in Fig \ref{dnn_framework}. As illustrated in Fig.~\ref{mdn}, UR-Net is a deep feedforward network with mixture density network layer as the final layer.

The input to UR-Net is of size $M Q$ and represents measured pairs of time-domain signals. We have $M$ sensor transmitter-receiver pair measurements (with sensors sparsely distributed across the space) and $Q$ discrete times points. We train the neural network with instances of our guided wave model in \eqref{lambwave}. We incorporate frequency-dependent wavenumber uncertainty into the simulation. We also add Gaussian noise to model sensor noise in the simulation. This is shown in the left-hand-side of Fig.~\ref{dnn_framework}. 

This paper considers active localization, where a signal is sent from a transmitter, interacts with the damage, and is then measured by the receiver  \cite{quazi1981overview}. In active localization, baseline subtraction is often first applied, removing the direct arrival and other reflections from the structure. Due to environmental variations, baseline subtraction is often imperfect \cite{dawson2015challenges}, but we do not consider these imperfections in this paper. We then standardize the data (zero mean, unit standard deviation) before inputting it into the network. 

As discussed in previous sub-section, the output of UR-Net (for d-dimensional localization) is of size $(2d+1) k$, where $k$ is the number of mixture components (corresponding to the number of potential damage locations) and $d$ is the output space dimension ($d$ = 2 in our case). Note that $k$ can be larger than the true number of damage locations. Unlikely component locations will have small likelihoods.  At inference time, UR-Net outputs the parameters conditioned on that one sample of test wave data. Predictive uncertainty at the output is represented by the estimated mixture component variance(s). The outputs can be visualized as a contour plot, shown at the right-hand-side of Fig.~\ref{dnn_framework}. The contour is the multi-modal distribution for the output conditioned on a particular input. The width of each Gaussian represents its predictive uncertainty.

\subsection{Matched Field Processing Comparison}
We compare UR-Net with matched field processing (MFP), a model-driven localization technique commonly used in underwater acoustics \cite{gingras1997electromagnetic}, seismology \cite{sidorovich1998two}, robotics \cite{argentieri2007broadband}, and radar \cite{yardibi2010source}. For MFP, there exists a query grid of dimensions $L \times W$. This grid would be superimposed on a structural plate or panel (such as on an aircraft). We assume the plate has point damage at some location, modeled as a point scatterer. 

Let $X(\omega_{q},r_{m})$ denote the signal for frequency $q$ and sensor pair $m$. Similarly, let $Z(\omega_{q},r_{m},p)$ be our mathematical model for damage at point $p$ in the grid. We define $Z(\omega_{q},r_{m},p)$ with the Lamb wave model described in \eqref{lambwave}. The correlation between the mathematical model and experimental data at point $p$ is denoted by 
\begin{equation}\label{mfp}
    b_{p} = \frac{|\sum_{m=1}^{M}\sum_{q=1}^{Q} X(\omega_{q},r_{m}) Z(\omega_{q},r_{m},p)^{*}|^{2}}{\sum_{m=1}^{M}\sum_{q=1}^{Q}|Z(\omega_{q},r_{m},p)|^{2}}
\end{equation}
where $(\cdot)^*$ denotes complex conjugate. The numerator of (\ref{mfp}) denotes the correlation between the experimental data and the mathematical model with the damage location assumed at a specific point $p$ in the grid. The denominator is a normalization factor. The location estimate by MFP is given by
\begin{equation}
      \widehat{b}_{p} = \arg\max_{p} b_{p} 
\end{equation}
\begin{figure*}[!t]
  \begin{subfigure}[h]{0.48\textwidth}
  \centering
  %\captionsetup{justification=centering}
    \includegraphics[width=\textwidth]{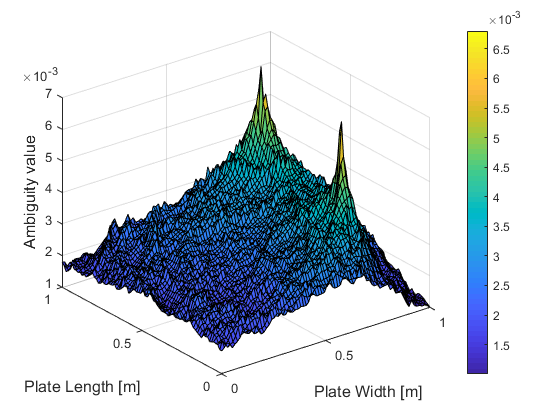}
    \caption{}
    \label{mfp_illustration}
  \end{subfigure}
  \hfill
  \begin{subfigure}[!t]{0.48\textwidth}
  \centering
  %\captionsetup{justification=centering}
    \includegraphics[width=\textwidth]{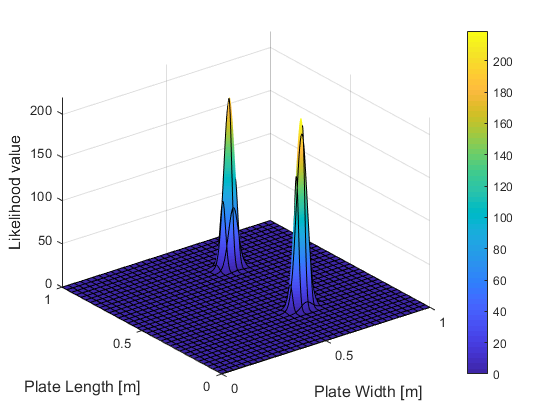}
    \caption{}
    \label{mdn_illustration}
  \end{subfigure}
  \caption{Representative localization contours produced by: (a) MFP and (b) UR-Net} 
\end{figure*}
where $p$ corresponds to some coordinate location. 

\indent Fig.~6 compares the MFP output $b_p$ with UR-Net output. MFP obtains a spread out and granular contour, representing the correlation between the physical wave model and experimental data. The contour obtained from UR-Net is a visualization of a mixture model, where large values correspond to locations with a high likelihood. Hence, uncertainty with UR-Net is encapsulated by the distributional parameters. With MFP, the uncertainty is defined by similarities between the data and the mathematical model. 

%We would like to remind the reader that the optimization procedure for training MDN involves maximizing the likelihood of a probability distribution. Consequently, the obtained distribution parameters place a high mass at points of high likelihood and vice-versa. This gives a sharp contour. This is in contrast with the output of MFP, where the output is a correlation value at every grid-point. In addition, our approach's output is a likelihood value for every grid-point. Hence, uncertainty can be encapsulated in a better manner by the distributional parameters compared to the data itself.

\section{Simulation Setup}

\subsection{Data Setup}

%\subsection{Simulation setup}
\begin{figure}[!h]
    \includegraphics[width= 9.0 cm]{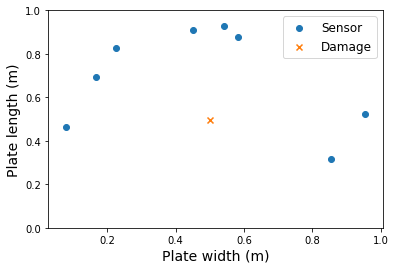}
    \caption{Illustration of the sensor placement for a single run of simulation setup}
    \label{simulation}
\end{figure}
We use Lamb wave simulations in (1) to generate both training and testing data. UR-Net and MFP are applied to the same dataset. We consider a region of a $1$~m by $1$~m plate. Damage(s) is modeled at random point(s), as shown in Fig.~\ref{simulation}. While there are existing methods for optimal sensor geometry \cite{OptSensorPlace}, \cite{OptimalGeometry}, we use random sensor placement to unbias the results based on sensor configuration. We simulate acoustic data using equation (1) with $m = 8$ sensors ($M = 28$ unique sensor pairs) placed at random locations on the plate. We use $Q = 1000$ equally spaced frequencies from $-500$ to $500$~kHz to simulate the waves. For UR-Net, we convert the input signal into the time-domain.

%We refer to the $Q \times M$ time-domain wave data matrix and the corresponding damage location(s) as one input-output sample.

% We use a sparse-array configuration for the sensors. Optimal methods for sensor array processing have been studied to a great extent in the past \cite{van2004optimum, yang2007different} \hl{What is the significance of stating this? It does not seem to have a purpose right now.}.

%\indent Consider $\mathbf{X}_b$ as the damage-free signal and $\mathbf{X}_s$ as the damage signal. $\widetilde{\mathbf{X}}_b$ is the modeled baseline signal. The environmental variations that affect wave propagation have to be incorporated during simulation. Due to the random nature of uncertainty, it follows that the baseline subtraction in our simulation is not perfect. The final simulation model is as follows
%\begin{equation}
%    \mathbf{X} = \mathbf{X}_b + \mathbf{X}_s - \widetilde{\mathbf{X}}_b + \epsilon
%\end{equation}
%where $\epsilon$ is additive white Gaussian noise used to model the uncertainty due to sensor noise.
%\hl{???? I do not think we need to include the discussion about the modeled basleine signal here since it is not part of this paper ????}
%\begin{equation}
%    \epsilon \sim \mathcal{N}(0, \sigma^{2})
%\end{equation}

The uncertainty due to environmental factors is modeled by introducing a random multiplicative effect $\alpha$ on the wavenumbers. The modified wavenumbers are defined as $\kappa(\omega) = \alpha \kappa(\omega)$. We statistically model $\alpha$ as
\begin{equation}
    \alpha \sim \mathcal{N}_{t}(1, 1, 1 - w_{distort}, 1 + w_{distort})
\end{equation}.
where $\mathcal{N}_{t}(\mu, \sigma^{2}, a, b)$ represents a truncated Gaussian distribution with mean $\mu$, variance $\sigma^{2}$, and truncated between $-\infty < a < b < \infty$. We vary the uncertainty in wavenumber by varying $w_{distort}$ (e.g., $w_{distort}$ = 0.15 implies that $\alpha$ is sampled from a distribution truncated between 0.85 and 1.15). We use a truncated Gaussian to simulate a wavenumber distortion that is as close to a realistic environmental scenario as possible.

\subsection{UR-Net implementation}
%\indent The time-domain wave data is pre-processed based on standard neural network practices. This includes standardizing it and flattening the matrix into a 1D vector to be used as input to UR-Net. 
UR-Net has a total of $4$~hidden layers, of which, $3$~are fully connected hidden layers. First hidden layer has $h_{1} = 600$~nodes, second hidden layer has $h_{2} = 300$~nodes, and third hidden layer has $h_{3} = 60$~nodes. The output layer is the MDN layer. The number of hidden layers and nodes in each layer is chosen after manual experimentation to maximize performance. This choice of hyperparameters gives the fastest convergence in terms of number of iterations. 

\indent We simulate $5000$ Monte Carlo samples (i.e., collections of measurements with different SNRs and $w_{distort}$ values) for training UR-Net. We use the Adam optimization algorithm \cite{kingma2014adam} for training the network. We also regularize UR-Net by using dropout \cite{srivastava2014dropout}, which is a common practice. Dropout randomly drops out nodes from the neural network during training time with a pre-specified probability. This potentially leads the network to learn weights in a robust manner by preventing overfitting.

\indent We use 3-fold cross-validation in our model training process. We observe that for the training phase, dropout probability has to be increased between $0.15$ to $0.25$ as the input uncertainty increases to avoid overfitting. To choose the right value of dropout probability, we monitor the training and validation loss trend. The framework is implemented using Keras framework \cite{chollet2015keras}. We report results and compare the performance of UR-Net and MFP on a separate test set consisting of $100$~samples. The test data and training data are configured with the same noise level, amount of wavenumber distortion ($w_{distort}$), and number of damage locations.

\subsection{MFP Setup}
As our simulation consists of a unit plate, we set unit grid dimensions for MFP (i.e L, W = 1). We define $N_{x} = N_{y} = 100$ equally spaced grid points on each axis (giving a total of 10000 grid points). For ease of localization with MFP, we divide the query grid into 4 equal sub-grids. We then simulate damage(s) in distinct sub-grids. The location(s) in distinct sub-grids (that have a damage) with the maximum correlation value are chosen as the MFP estimate(s). For comparison, we use the same test data as used for UR-Net.

\section{Results and Discussion}
During inference time, every predicted component mean $\mu_{k}$ for UR-Net is assigned to the closest true damage location. The performance of UR-Net is quantified on the test set using an average localization error (ALE) metric, defined as:
\begin{equation}\label{ale}
    \textrm{ALE} = \frac{1}{T}\sum_{i = 1}^{T}\frac{1}{K_{i}}\sum_{j=1}^{K_{i}}\sqrt{(x_{ij} - \overline{x}_{ij})^{2} + (y_{ij} - \overline{y}_{ij})^{2}}
\end{equation}
Here $T$ is the total number of samples in the data-set and $K_{i}$ is the number of damages in the $i^{th}$ sample. Further, $(x_{ij},y_{ij})$, $(\overline{x}_{ij},\overline{y}_{ij})$ are the actual and predicted damage locations respectively for the $j^{th}$ damage in the $i^{th}$ sample. ALE is the Euclidean distance between the actual and predicted damage location, averaged across each sample of the dataset. A smaller value of ALE indicates that the localization prediction is more accurate. For all of the further experiments, we choose a GMM with 3 components unless mentioned otherwise. %The number of mixture components is a user choice. 
For MFP, the ALE is computed by plugging in the estimated damage location(s) in \eqref{ale}.

\subsection{Performance comparison at varying levels of noise}
\begin{figure}
    \includegraphics[width= 9.0 cm]{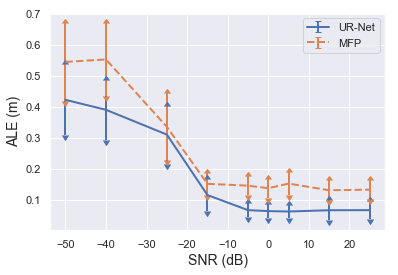}
    \caption{Comparing ALE of UR-Net and MFP}
    \label{snr_variation}
\end{figure}

\indent Fig.~\ref{snr_variation} shows the ALE comparison of UR-Net and MFP at different sensor noise levels for a simulation setup with a maximum of 2 damages. In this illustration, the error bars represent the standard deviation in the prediction errors. In this simulation, wavenumber distortion ($w_{distort}$) has been set to 0.15. Sensor noise is modeled as additive, white Gaussian noise (AWGN). Here, signal-to-noise ratio (SNR) is defined in as
\begin{equation}
     \textrm{SNR}_{dB} = 10\log_{10}\left(\frac{\textrm{Signal}_{\textrm{power}}}{\textrm{Noise}_{\textrm{power}}}\right)
\end{equation}
where $\textrm{Signal}_{\textrm{power}}$ is the power of the baseline-subtracted signal. 
 
\indent We observe that UR-Net has a substantially better performance than MFP. We also observe a trend of decreasing ALE as the signal-to-noise ratio (SNR) goes up. UR-Net has an error of 0.0625 m at 25 dB as compared to 0.1425 m for MFP. While the error at -50 dB is 0.43 m and 0.54 m, respectively, for the two approaches. Note that the error convergences to approximately $0.5$~m, one-half the length and width of plate. Based on a Monte Carlo analysis, the expected error of randomly sampling true and predicted locations with a uniform distribution would be approximately~$0.52$~m. 

\subsection{Performance comparison at varying levels of uncertainty}
\begin{figure}[!t]
    \includegraphics[width= 8.8 cm]{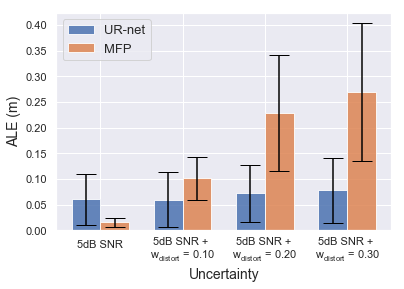}
    \caption{Performance comparison of MFP and and UR-Net for varying levels of uncertainty}
    \label{uncertainty_variation}
\end{figure}

\indent Fig.~\ref{uncertainty_variation} illustrates the performance of UR-Net and MFP at varying levels of uncertainty in the simulation setup. The leftmost bar represents data with 5 dB SNR. The next three bars represent scenarios with increasing levels of wavenumber uncertainty and 5 dB SNR. In the case with only noise, MFP has optimal performance ($\approx 0.0152$~m), while performance of UR-Net is comparably inferior ($\approx 0.06$~m). This is reasonable since MFP is derived from the matched filter, an optimal detector in white Gaussian noise.

\indent The next three bars represent cases where we have wavenumber distortion and $5$~dB SNR (both epistemic and aleatoric uncertainty). We observe that while the performance of UR-Net sees only a small decrease with varying uncertainty levels, the performance of MFP decreases consistently with increasing uncertainty level. MFP cannot handle uncertainty in wavenumber since the underlying model of the MFP is fixed. In contrast, since our approach learns from the available data to infer the conditional output distribution, it is more robust to uncertainty. 

\begin{figure}[!t]
    \includegraphics[width= 9.0 cm]{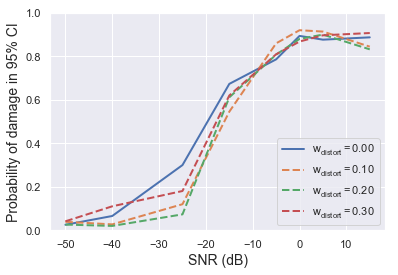}
    \caption{Probability of damage in 95\% confidence interval}
    \label{ci95}
\end{figure}

\indent Fig.~\ref{ci95} compares the probability of true damage being within a $95\%$ confidence interval of a component mean for different values of wavenumber distortion ($w_{distort}$). A high probability of damage being in the $95\%$ confidence interval tells us that the mean and variance parameters of the inferred conditional distribution are able to reliably model the uncertainty in the conditional distribution of the damage location given the input data. 

\indent We observe that the curves for all values of $w_{distort}$ are closely overlapping in the SNR range of $0$ to $15$~dB.This trend implies that even with increasing levels of uncertainty in wavenumber, UR-Net is able to infer appropriate parameters for the conditional output distribution. With the evidence of a trend in this figure and the trend of nearly constant performance of UR-Net in the previous figure (Fig.~\ref{uncertainty_variation}), we observe that the framework performance is robust to uncertainty in velocity (i.e., $w_{distort}$).

\subsection{Representing predictive uncertainty with UR-net}
\begin{figure}[!t]
    \includegraphics[width= 9.0 cm]{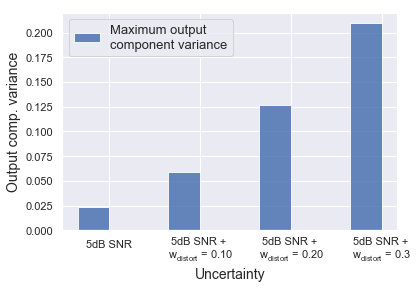}
    \caption{Variance as a measure of predictive uncertainty in UR-Net}
    \label{var_upperbound}
\end{figure}
\begin{figure}[!t]
    \includegraphics[width= 9.0 cm]{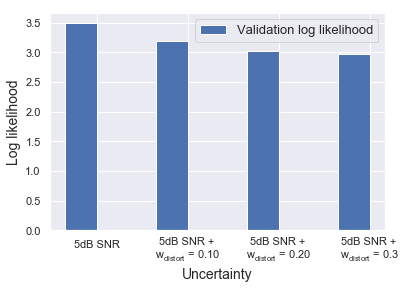}
    \caption{Log likelihood for UR-Net in presence of varying uncertainty}
    \label{loglikelihood}
\end{figure}

\indent While robustness to uncertainty is one goal, another goal is to represent the predictive uncertainty. We explore the use of component variance parameter output by UR-Net in (9) to represent predictive uncertainty. Fig.~\ref{var_upperbound} illustrates the maximum value of component variance (worst value) obtained from UR-Net during inference time with varying levels of uncertainty. We observe a trend of increasing predictive variance as uncertainty ($w_{distort}$) increases. The leftmost bar representing a noise-only scenario has the lowest predictive variance ($\approx 0.023$) while the rightmost bar with $5$~dB~SNR and $w_{distort} = 0.3$ has the maximum predictive variance ($ \approx 0.209$). The component variance parameter outputted by the UR-Net thus can act as an efficient way to represent uncertainty that propagates from input to output.

\indent Fig.~\ref{loglikelihood} illustrates the average log-likelihood for the validation data set as outputted by the UR-Net. This comparison is in the presence of varying levels of uncertainty. We  observe a trend of decreasing likelihood with increasing level of uncertainty. With only noise, the log-likelihood is $\approx$ 3.5 and drops to $\approx$ 3.0 when we have both noise and $w_{distort} = 0.3$. Here, likelihood can be thought of as a measure of ``goodness of fit'' for the inferred model. %It can also serve as a measure of uncertainty by comparing it with the likelihood value obtained after training a network on ideal data \hl{Not sure what you mean by this}.

\subsection{Performance comparison with varying number of damages}
\begin{figure}[!t]
    \includegraphics[width= 9.0 cm]{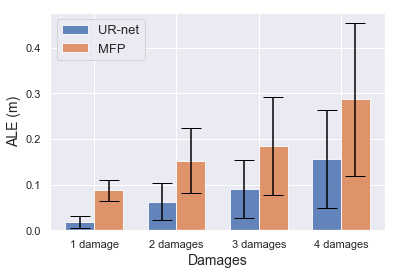}
    \caption{Comparing ALE of UR-Net and MFP with varying number of damages}
    \label{damage_variation}
\end{figure}

%\begin{figure}[!h]
%    \includegraphics[width= 9.0 %cm]{IEEE_tran/ci95_damagevariation.png}
%    \caption{Probability of damage in 95\% confidence interval}
%    \label{ci95_damage}
%\end{figure}

\indent Fig.~\ref{damage_variation} shows the comparison between UR-Net and MFP with a varying number of damages in the simulation setup. In the simulation setup, we have wavenumber distortion ($w_{distort} = 0.15$) and $5$~dB sensor noise.  For UR-Net, we use one more mixture component than number of damages in the simulation setup. 

\indent The performance of UR-Net is superior to MFP for all number of damages investigated here. The localization error (ALE) for UR-Net is $\approx 0.0187$~m and $\approx 0.1563$~m for 1 and 4 damages respectively. Similarly, the corresponding error values for MFP are $\approx 0.0875$~m and $\approx 0.2873$~m respectively. While the performance of both approaches drops down with an increasing number of damages, MFP performance decreases at a faster rate as compared to UR-Net. 

%Fig.~\ref{ci95_damage} compares the probability of damage being in $95\%$ confidence interval for different numbers of damages. We observe that the probability is substantially higher for 1 damage. The performance trend is closely similar for 2,3, and 4 damages. The overall trend can be attributed to the fact that the dual task of both isolating and localizing damage becomes tougher as the number of damages goes on increasing. 

\section{Conclusion}
We proposed an uncertainty-aware DNN based framework, UR-Net, for damage localization with guided waves. The motivation for our framework was the need to assess and represent the uncertainty due to environmental variations. Our framework outputs parameters of an appropriately chosen mixture model. The outputted mixture component means are estimates for the damage locations and the component variances are a measure of the predictive uncertainty.

\indent We showed the results of various simulation runs that compare the localization performance of UR-Net with MFP. Our framework shows better performance than MFP in presence of aleatoric and epistemic uncertainty. We also analyze the effect of increasing levels of uncertainty on the performance of our approach. Finally, we studied the effect of increasing the number of damages in the performance of both MFP and UR-Net.

\indent We showed that UR-Net has two advantages over MFP: robustness to uncertainty due to environmental variations and being able to represent damage location as a probability distribution for a multi-damage scenario. Representing damage location as a probability distribution offers us a principled way of representing uncertainty through confidence intervals and / or variance.

\indent In the future, the approach can be extended for random structure geometries. Further work will also include analysis of the resolution achievable by this framework. This can serve as a metric of comparison with traditional localization algorithms. In the future, there is also a potential to explore ways for estimating damage intensity using this framework.

% use section* for acknowledgment
\section*{Acknowledgment}
This research is supported by the Air Force Office of Scientific Research under award number FA9550-17-1-0126 and by the National Science Foundation under award number EECS-1839704.

%The authors would like to thank... \hl{Need to acknowledge NSF and AFOSR grant}

% Can use something like this to put references on a page
% by themselves when using endfloat and the captionsoff option.
\ifCLASSOPTIONcaptionsoff
  \newpage
\fi

% trigger a \newpage just before the given reference
% number - used to balance the columns on the last page
% adjust value as needed - may need to be readjusted if
% the document is modified later
%\IEEEtriggeratref{8}
% The "triggered" command can be changed if desired:
%\IEEEtriggercmd{\enlargethispage{-5in}}

% references section

% can use a bibliography generated by BibTeX as a .bbl file
% BibTeX documentation can be easily obtained at:
% http://mirror.ctan.org/biblio/bibtex/contrib/doc/
% The IEEEtran BibTeX style support page is at:
% http://www.michaelshell.org/tex/ieeetran/bibtex/
%\bibliographystyle{IEEEtran}
% argument is your BibTeX string definitions and bibliography database(s)
%\bibliography{IEEEabrv,../bib/paper}
%
% <OR> manually copy in the resultant .bbl file
% set second argument of \begin to the number of references
% (used to reserve space for the reference number labels box)
\bibliographystyle{IEEEtran}
\bibliography{bare_jrnl}

% Generated by IEEEtran.bst, version: 1.14 (2015/08/26)
\begin{thebibliography}{10}
\providecommand{\url}[1]{#1}
\csname url@samestyle\endcsname
\providecommand{\newblock}{\relax}
\providecommand{\bibinfo}[2]{#2}
\providecommand{\BIBentrySTDinterwordspacing}{\spaceskip=0pt\relax}
\providecommand{\BIBentryALTinterwordstretchfactor}{4}
\providecommand{\BIBentryALTinterwordspacing}{\spaceskip=\fontdimen2\font plus
\BIBentryALTinterwordstretchfactor\fontdimen3\font minus
  \fontdimen4\font\relax}
\providecommand{\BIBforeignlanguage}[2]{{%
\expandafter\ifx\csname l@#1\endcsname\relax
\typeout{** WARNING: IEEEtran.bst: No hyphenation pattern has been}%
\typeout{** loaded for the language `#1'. Using the pattern for}%
\typeout{** the default language instead.}%
\else
\language=\csname l@#1\endcsname
\fi
#2}}
\providecommand{\BIBdecl}{\relax}
\BIBdecl

\bibitem{underwatersonar}
P.~A. {Forero}, P.~A. {Baxley}, and L.~{Straatemeier}, ``A multitask learning
  framework for broadband source-location mapping using passive sonar,''
  \emph{IEEE Transactions on Signal Processing}, vol.~63, no.~14, pp.
  3599--3614, 2015.

\bibitem{sidorovich1998two}
D.~V. Sidorovich and A.~B. Gershman, ``Two-dimensional wideband interpolated
  root-music applied to measured seismic data,'' \emph{IEEE Transactions on
  Signal Processing}, vol.~46, no.~8, pp. 2263--2267, 1998.

\bibitem{safavi2017localization}
S.~Safavi and U.~A. Khan, ``Localization in mobile networks via virtual convex
  hulls,'' \emph{IEEE Transactions on Signal and Information Processing over
  Networks}, vol.~4, no.~1, pp. 188--201, 2017.

\bibitem{AdaptiveRadar}
D.~{Orlando} and G.~{Ricci}, ``Adaptive radar detection and localization of a
  point-like target,'' \emph{IEEE Transactions on Signal Processing}, vol.~59,
  no.~9, pp. 4086--4096, 2011.

\bibitem{alpay2000model}
M.~E. Alpay and M.~H. Shor, ``Model-based solution techniques for the source
  localization problem,'' \emph{IEEE Transactions on Control Systems
  Technology}, vol.~8, no.~6, pp. 895--904, 2000.

\bibitem{BeamformingLocalization}
L.~{Kumar} and R.~M. {Hegde}, ``Near-field acoustic source localization and
  beamforming in spherical harmonics domain,'' \emph{IEEE Transactions on
  Signal Processing}, vol.~64, no.~13, pp. 3351--3361, 2016.

\bibitem{zuo2018subspace}
W.~Zuo, J.~Xin, N.~Zheng, and A.~Sano, ``Subspace-based localization of
  far-field and near-field signals without eigendecomposition,'' \emph{IEEE
  Transactions on Signal Processing}, vol.~66, no.~17, pp. 4461--4476, 2018.

\bibitem{le2016closed}
T.-K. Le and N.~Ono, ``Closed-form and near closed-form solutions for toa-based
  joint source and sensor localization,'' \emph{IEEE Transactions on Signal
  Processing}, vol.~64, no.~18, pp. 4751--4766, 2016.

\bibitem{ko2005technology}
J.~Ko and Y.~Q. Ni, ``Technology developments in structural health monitoring
  of large-scale bridges,'' \emph{Engineering structures}, vol.~27, no.~12, pp.
  1715--1725, 2005.

\bibitem{rose2009successes}
J.~L. Rose, ``Successes and challenges in ultrasonic guided waves for ndt and
  shm,'' in \emph{Proc. of the National Seminar \& Exhibition on
  Non-Destructive Evaluation}, 2009, pp. 10--12.

\bibitem{mitra2016guided}
M.~Mitra and S.~Gopalakrishnan, ``Guided wave based structural health
  monitoring: A review,'' \emph{Smart Materials and Structures}, vol.~25,
  no.~5, p. 053001, 2016.

\bibitem{friswell2007damage}
M.~I. Friswell, ``Damage identification using inverse methods,''
  \emph{Philosophical Transactions of the Royal Society A: Mathematical,
  Physical and Engineering Sciences}, vol. 365, no. 1851, pp. 393--410, 2007.

\bibitem{giannakis1990signal}
G.~B. Giannakis and M.~K. Tsatsanis, ``Signal detection and classification
  using matched filtering and higher order statistics,'' \emph{IEEE
  Transactions on Acoustics, Speech, and Signal Processing}, vol.~38, no.~7,
  pp. 1284--1296, 1990.

\bibitem{harley2014data}
J.~B. Harley and J.~M. Moura, ``Data-driven matched field processing for lamb
  wave structural health monitoring,'' \emph{The Journal of the Acoustical
  Society of America}, vol. 135, no.~3, pp. 1231--1244, 2014.

\bibitem{farrar2007introduction}
C.~R. Farrar and K.~Worden, ``An introduction to structural health
  monitoring,'' \emph{Philosophical Transactions of the Royal Society A:
  Mathematical, Physical and Engineering Sciences}, vol. 365, no. 1851, pp.
  303--315, 2007.

\bibitem{lorenzoni2016uncertainty}
F.~Lorenzoni, F.~Casarin, M.~Caldon, K.~Islami, and C.~Modena, ``Uncertainty
  quantification in structural health monitoring: Applications on cultural
  heritage buildings,'' \emph{Mechanical Systems and Signal Processing},
  vol.~66, pp. 268--281, 2016.

\bibitem{sankararaman2011uncertainty}
S.~Sankararaman and S.~Mahadevan, ``Uncertainty quantification in structural
  damage diagnosis,'' \emph{Structural Control and Health Monitoring}, vol.~18,
  no.~8, pp. 807--824, 2011.

\bibitem{EllipticLocalization}
L.~{Rui} and K.~C. {Ho}, ``Elliptic localization: Performance study and optimum
  receiver placement,'' \emph{IEEE Transactions on Signal Processing}, vol.~62,
  no.~18, pp. 4673--4688, 2014.

\bibitem{RecLocnError}
K.~C. {Ho}, X.~{Lu}, and L.~{Kovavisaruch}, ``Source localization using tdoa
  and fdoa measurements in the presence of receiver location errors: Analysis
  and solution,'' \emph{IEEE Transactions on Signal Processing}, vol.~55,
  no.~2, pp. 684--696, 2007.

\bibitem{feuillade1989environmental}
C.~Feuillade, D.~Del~Balzo, and M.~M. Rowe, ``Environmental mismatch in
  shallow-water matched-field processing: Geoacoustic parameter variability,''
  \emph{The Journal of the Acoustical Society of America}, vol.~85, no.~6, pp.
  2354--2364, 1989.

\bibitem{StochasticMFP}
Y.~{Le Gall}, F.~{Socheleau}, and J.~{Bonnel}, ``Matched-field processing
  performance under the stochastic and deterministic signal models,''
  \emph{IEEE Transactions on Signal Processing}, vol.~62, no.~22, pp.
  5825--5838, 2014.

\bibitem{Temperature}
G.~{Konstantinidis}, P.~D. {Wilcox}, and B.~W. {Drinkwater}, ``An investigation
  into the temperature stability of a guided wave structural health monitoring
  system using permanently attached sensors,'' \emph{IEEE Sensors Journal},
  vol.~7, no.~5, pp. 905--912, 2007.

\bibitem{raghavan2008effects}
A.~Raghavan and C.~E. Cesnik, ``Effects of elevated temperature on guided-wave
  structural health monitoring,'' \emph{Journal of Intelligent Material Systems
  and Structures}, vol.~19, no.~12, pp. 1383--1398, 2008.

\bibitem{eybpoosh2014investigation}
M.~Eybpoosh, M.~Berg{\'e}s, and H.~Noh, ``Investigation on the effects of
  environmental and operational conditions (eoc) on diffuse-field ultrasonic
  guided-waves in pipes,'' in \emph{Computing in Civil and Building
  Engineering}, 2014, pp. 1198--1205.

\bibitem{harley2012scale}
J.~B. Harley and J.~M. Moura, ``Scale transform signal processing for optimal
  ultrasonic temperature compensation,'' \emph{IEEE Transactions on
  Ultrasonics, Ferroelectrics, and Frequency Control}, vol.~59, no.~10, pp.
  2226--2236, 2012.

\bibitem{InterferenceManagement}
H.~{Sun}, X.~{Chen}, Q.~{Shi}, M.~{Hong}, X.~{Fu}, and N.~D. {Sidiropoulos},
  ``Learning to optimize: Training deep neural networks for interference
  management,'' \emph{IEEE Transactions on Signal Processing}, vol.~66, no.~20,
  pp. 5438--5453, 2018.

\bibitem{MIMODeepLearning}
H.~{He}, C.~{Wen}, S.~{Jin}, and G.~Y. {Li}, ``Model-driven deep learning for
  mimo detection,'' \emph{IEEE Transactions on Signal Processing}, vol.~68, pp.
  1702--1715, 2020.

\bibitem{ResourceAlloc}
M.~{Eisen}, C.~{Zhang}, L.~F.~O. {Chamon}, D.~D. {Lee}, and A.~{Ribeiro},
  ``Learning optimal resource allocations in wireless systems,'' \emph{IEEE
  Transactions on Signal Processing}, vol.~67, no.~10, pp. 2775--2790, 2019.

\bibitem{niu2017source}
H.~Niu, E.~Reeves, and P.~Gerstoft, ``Source localization in an ocean waveguide
  using supervised machine learning,'' \emph{The Journal of the Acoustical
  Society of America}, vol. 142, no.~3, pp. 1176--1188, 2017.

\bibitem{yalta2017sound}
N.~Yalta, K.~Nakadai, and T.~Ogata, ``Sound source localization using deep
  learning models,'' \emph{Journal of Robotics and Mechatronics}, vol.~29,
  no.~1, pp. 37--48, 2017.

\bibitem{tompson2015efficient}
J.~Tompson, R.~Goroshin, A.~Jain, Y.~LeCun, and C.~Bregler, ``Efficient object
  localization using convolutional networks,'' in \emph{Proc. of the IEEE
  Conference on Computer Vision and Pattern Recognition}, 2015, pp. 648--656.

\bibitem{bengio2012deep}
Y.~Bengio, ``Deep learning of representations for unsupervised and transfer
  learning,'' in \emph{Proc. of ICML Workshop on Unsupervised and Transfer
  Learning}, 2012, pp. 17--36.

\bibitem{harley2017managing}
J.~B. Harley, C.~Liu, I.~J. Oppenheim, and J.~M. Moura, ``Managing complexity,
  uncertainty, and variability in guided wave structural health monitoring,''
  \emph{SICE Journal of Control, Measurement, and System Integration}, vol.~10,
  no.~5, pp. 325--336, 2017.

\bibitem{tian2014lamb}
Z.~Tian and L.~Yu, ``Lamb wave frequency--wavenumber analysis and
  decomposition,'' \emph{Journal of Intelligent Material Systems and
  Structures}, vol.~25, no.~9, pp. 1107--1123, 2014.

\bibitem{alleyne1992interaction}
D.~N. Alleyne and P.~Cawley, ``The interaction of lamb waves with defects,''
  \emph{IEEE Transactions on Ultrasonics, Ferroelectrics, and Frequency
  Control}, vol.~39, no.~3, pp. 381--397, 1992.

\bibitem{chowdhary_dupuis_2013}
K.~Chowdhary and P.~Dupuis, ``Distinguishing and integrating aleatoric and
  epistemic variation in uncertainty quantification,'' \emph{ESAIM:
  Mathematical Modelling and Numerical Analysis}, vol.~47, no.~3, p. 635–662,
  2013.

\bibitem{weaver1981dispersion}
R.~L. Weaver and Y.-H. Pao, ``Dispersion relations for linear wave propagation
  in homogeneous and inhomogeneous media,'' \emph{Journal of Mathematical
  Physics}, vol.~22, no.~9, pp. 1909--1918, 1981.

\bibitem{bishop1994mixture}
C.~M. Bishop, ``Mixture density networks,'' \emph{Technical Report NCRG/94/004,
  Neural Computing Research Group}, 1994.

\bibitem{choi2018uncertainty}
S.~Choi, K.~Lee, S.~Lim, and S.~Oh, ``Uncertainty-aware learning from
  demonstration using mixture density networks with sampling-free variance
  modeling,'' in \emph{Proc. of the IEEE International Conference on Robotics
  and Automation}.\hskip 1em plus 0.5em minus 0.4em\relax IEEE, 2018, pp.
  6915--6922.

\bibitem{zen2014deep}
H.~Zen and A.~Senior, ``Deep mixture density networks for acoustic modeling in
  statistical parametric speech synthesis,'' in \emph{Proc. of the IEEE
  International Conference on Acoustics, Speech and Signal Processing}.\hskip
  1em plus 0.5em minus 0.4em\relax IEEE, 2014, pp. 3844--3848.

\bibitem{wang2016gating}
W.~Wang, S.~Xu, and B.~Xu, ``Gating recurrent mixture density networks for
  acoustic modeling in statistical parametric speech synthesis,'' in
  \emph{Proc. of the IEEE International Conference on Acoustics, Speech and
  Signal Processing}.\hskip 1em plus 0.5em minus 0.4em\relax IEEE, 2016, pp.
  5520--5524.

\bibitem{wang2017autoregressive}
X.~Wang, S.~Takaki, and J.~Yamagishi, ``An autoregressive recurrent mixture
  density network for parametric speech synthesis,'' in \emph{Proc. of the IEEE
  International Conference on Acoustics, Speech and Signal Processing}.\hskip
  1em plus 0.5em minus 0.4em\relax IEEE, 2017, pp. 4895--4899.

\bibitem{bernardo2009bayesian}
J.~M. Bernardo and A.~F. Smith, \emph{Bayesian theory}.\hskip 1em plus 0.5em
  minus 0.4em\relax John Wiley \& Sons, 2009, vol. 405.

\bibitem{neal2012bayesian}
R.~M. Neal, \emph{Bayesian learning for neural networks}.\hskip 1em plus 0.5em
  minus 0.4em\relax Springer Science \& Business Media, 2012, vol. 118.

\bibitem{gal2016dropout}
Y.~Gal and Z.~Ghahramani, ``Dropout as a bayesian approximation: Representing
  model uncertainty in deep learning,'' in \emph{Proc. of the International
  Conference on Machine Learning}, 2016, pp. 1050--1059.

\bibitem{mclachlan2004finite}
G.~J. McLachlan and D.~Peel, \emph{Finite mixture models}.\hskip 1em plus 0.5em
  minus 0.4em\relax John Wiley \& Sons, 2004.

\bibitem{kostantinos2000gaussian}
N.~Kostantinos, ``Gaussian mixtures and their applications to signal
  processing,'' \emph{Advanced Signal Processing Handbook: Theory and
  Implementation for Radar, Sonar, and Medical Imaging Real Time Systems}, pp.
  3--1, 2000.

\bibitem{mdn}
C.~Martin, ``Keras mixture density network layer,''
  \url{https://github.com/cpmpercussion/keras-mdn-layer}, 2019.

\bibitem{quazi1981overview}
A.~Quazi, ``An overview on the time delay estimate in active and passive
  systems for target localization,'' \emph{IEEE Transactions on Acoustics,
  Speech, and Signal Processing}, vol.~29, no.~3, pp. 527--533, 1981.

\bibitem{dawson2015challenges}
A.~J. Dawson, J.~E. Michaels, and T.~E. Michaels, ``Challenges in the
  separation and analysis of scattered waves in angle-beam wavefield data,'' in
  \emph{Proc. of the Review of Quantitative Nondestructive Evaluation}, vol.
  1650, no.~1.\hskip 1em plus 0.5em minus 0.4em\relax American Institute of
  Physics, 2015, pp. 827--834.

\bibitem{gingras1997electromagnetic}
D.~F. Gingras, P.~Gerstoft, N.~L. Gerr, and C.~Mecklenbrauker,
  ``Electromagnetic matched field processing for source localization,'' in
  \emph{Proc. of the IEEE International Conference on Acoustics, Speech, and
  Signal Processing}, vol.~1.\hskip 1em plus 0.5em minus 0.4em\relax IEEE,
  1997, pp. 479--482.

\bibitem{argentieri2007broadband}
S.~Argentieri and P.~Danes, ``Broadband variations of the music high-resolution
  method for sound source localization in robotics,'' in \emph{Proc. of the
  IEEE/RSJ International Conference on Intelligent Robots and Systems}.\hskip
  1em plus 0.5em minus 0.4em\relax IEEE, 2007, pp. 2009--2014.

\bibitem{yardibi2010source}
T.~Yardibi, J.~Li, P.~Stoica, M.~Xue, and A.~B. Baggeroer, ``Source
  localization and sensing: A nonparametric iterative adaptive approach based
  on weighted least squares,'' \emph{IEEE Transactions on Aerospace and
  Electronic Systems}, vol.~46, no.~1, pp. 425--443, 2010.

\bibitem{OptSensorPlace}
M.~{Sadeghi}, F.~{Behnia}, and R.~{Amiri}, ``Optimal sensor placement for 2-d
  range-only target localization in constrained sensor geometry,'' \emph{IEEE
  Transactions on Signal Processing}, vol.~68, pp. 2316--2327, 2020.

\bibitem{OptimalGeometry}
N.~H. {Nguyen} and K.~{Doğançay}, ``Optimal geometry analysis for multistatic
  toa localization,'' \emph{IEEE Transactions on Signal Processing}, vol.~64,
  no.~16, pp. 4180--4193, 2016.

\bibitem{kingma2014adam}
D.~P. Kingma and J.~Ba, ``Adam: A method for stochastic optimization,''
  \emph{arXiv preprint arXiv:1412.6980}, 2014.

\bibitem{srivastava2014dropout}
N.~Srivastava, G.~Hinton, A.~Krizhevsky, I.~Sutskever, and R.~Salakhutdinov,
  ``Dropout: A simple way to prevent neural networks from overfitting,''
  \emph{The Journal of Machine Learning Research}, vol.~15, no.~1, pp.
  1929--1958, 2014.

\bibitem{chollet2015keras}
F.~Chollet \emph{et~al.}, ``Keras,'' \url{https://keras.io}, 2015.

\end{thebibliography}

%\begin{thebibliography}{sampbib}

%\bibitem{IEEEhowto:kopka}
%H.~Kopka and P.~W. Daly, \emph{A Guide to \LaTeX}, %3rd~ed.\hskip 1em plus
%  0.5em minus 0.4em\relax Harlow, England: Addison-Wesley, 1999.

%\end{thebibliography}

% biography section
% 
% If you have an EPS/PDF photo (graphicx package needed) extra braces are
% needed around the contents of the optional argument to biography to prevent
% the LaTeX parser from getting confused when it sees the complicated
% \includegraphics command within an optional argument. (You could create
% your own custom macro containing the \includegraphics command to make things
% simpler here.)

\begin{IEEEbiography}[{\includegraphics[width=1in,height=1.25in,clip,keepaspectratio]{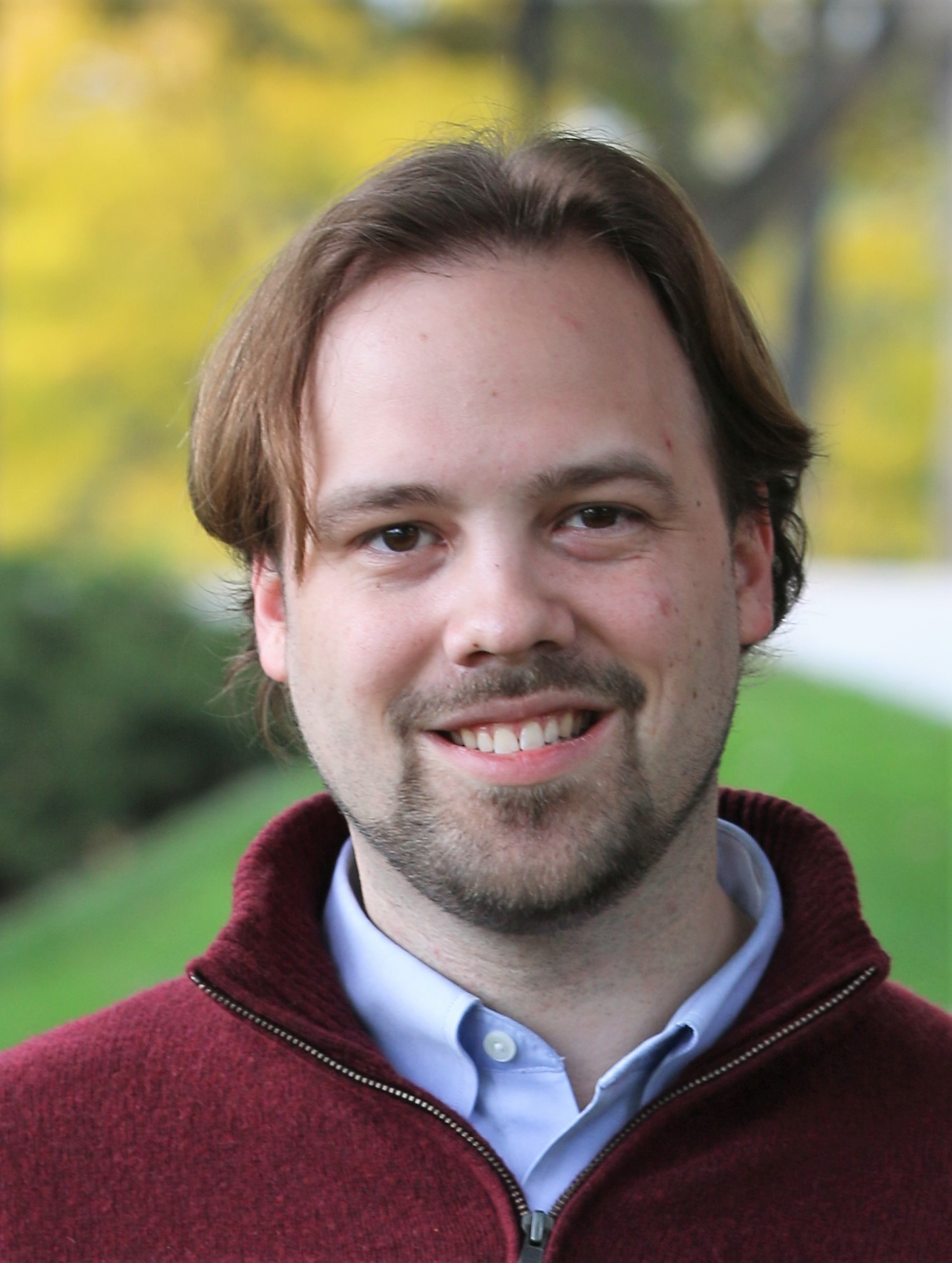}}]
Joel B. Harley (S'05-M'14) received his B.S. degree in Electrical Engineering from Tufts University in Medford, MA, USA. He received his M.S. and Ph.D. degrees in Electrical and Computer Engineering from Carnegie Mellon University in Pittsburgh, PA, USA in 2011 and 2014, respectively.

In 2018, he joined the University of Florida, where he is currently an assistant professor in the Department of Electrical and Computer Engineering. Previously, he was an assistant professor in the Department of Electrical and Computer Engineering at the University of Utah. His research interests include integrating novel signal processing, machine learning, and data science methods for the analysis of waves and time-series data.

Dr. Harley's awards and honors include  2020 IEEE Ultrasonics, Ferroelectrics, and Frequency Control Society Star Ambassador Award, a 2020 and 2018 Air Force Summer Faculty Fellowship, a 2017 Air Force Young Investigator Award, a 2014 Carnegie Mellon A. G. Jordan Award (for academic excellence and exceptional service to the community). He has published more than 90 technical journal and conference papers, including four best student papers. He is a student representative advisor for the IEEE Ultrasonics, Ferroelectrics, and Frequency Control Society, a member of the IEEE Signal Processing Society, and a member of the Acoustical Society of America.
\end{IEEEbiography}

\begin{IEEEbiography}[{\includegraphics[width=1.0in,height=1.25in, clip, keepaspectratio]{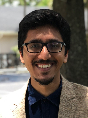}}]
Ishan D. Khurjekar received his B.S degree in Electronics and Instrumentation Engineering from BITS Pilani, Goa, India. He received his M.S degree in Electrical and Computer Engineering from Texas A\&M University in College Station, TX, USA in 2018. 

In 2018, he joined University of Florida, where he is currently pursuing a Ph.D degree in Electrical and Computer Engineering, advised by Dr Joel B. Harley. His research interests include signal processing, source localization, machine learning and its applications in inverse problem solving, probabilistic deep learning, probabilistic graphical models, uncertainty quantification. 

\end{IEEEbiography}
\vfill
% insert where needed to balance the two columns on the last page with
% biographies
%\newpage

% You can push biographies down or up by placing
% a \vfill before or after them. The appropriate
% use of \vfill depends on what kind of text is
% on the last page and whether or not the columns
% are being equalized.

%\vfill

% Can be used to pull up biographies so that the bottom of the last one
% is flush with the other column.
%\enlargethispage{-5in}

% that's all folks
\end{document}